\documentclass{svproc}

\usepackage[numbers]{natbib}
\usepackage[utf8]{inputenc}

\usepackage{todonotes}

\usepackage{placeins}

\usepackage{url}

\usepackage{hyperref}

\usepackage{url}

\usepackage{booktabs}

\usepackage{diagbox}

\title{A Quantitative Exploration of the 9-Factor Theory: Distribution of Leadership Roles between Scrum Master and Agile Team}

\author{Simone V. Spiegler \inst{1,2} \and Daniel Graziotin \inst{1} \and Christoph Heinecke \inst{3} \and Stefan Wagner \inst{1}}

	\authorrunning{S. V. Spiegler et al.}
	
	\institute{Institute of Software Technology, University of Stuttgart, Germany\\ \email{\{simone.spiegler,daniel.graziotin,stefan.wagner\}@iste.uni-stuttgart.de}
		\and Robert Bosch Automotive Steering GmbH, Schw\"abisch Gm\"und, Germany\\
		\and Robert Bosch GmbH, Stuttgart, Germany\\
		\email{Christoph.Heinecke@bosch.com}}
	
	\titlerunning{Distribution of Leadership Roles}

\begin{document}

\maketitle

\begin{abstract}

A number of qualitative studies find that team leadership is one essential success factor for evolving into a mature agile team. One such qualitative study suggests the 9-Factor Theory of Scrum Master roles, which claims that the Scrum Master performs a set of 9 leadership roles which are transferred to the team over time \cite{spiegler2019}.

We aimed at conducting a quantitative exploration that examines the presence and change of the  9-Factory Theory in relation to team maturity.

We conducted an online survey with 67 individuals at the conglomerate Robert Bosch GmbH. Descriptive statistics reveal that the Scrum Master and the agile team score differently on the 9 factors and that the Scrum Master role is most often distributed in teams that had been working between 3 and 5 months in an agile manner. Yet, we also find that the leadership roles predominantly remain with one dedicated Scrum Master.

Based on our results we suggest to group the 9-Factor Theory into three clusters: the Scrum Master is rather linked to psychological team factors (1), while the team tends to be linked to rather product-related factors (2). Organizational factors (3) are less often present.

Our practical implications suggest an extension of the Scrum Master description. Furthermore, our study lays groundwork for future quantitative testing of leadership in agile teams.

\end{abstract}

\section{Introduction}

Even though an increasing number of organizations aim at implementing agile teams, how to do so is not yet entirely clear \cite{nerur2005challenges, Moe2010}. Especially rather bureaucratic companies seem to struggle in their agile transformation \cite{nerur2005challenges}.

Fitting leadership behavior is found to be one key success factor for evolving into an agile team \cite{gren2019agile}. The agile way of working suggests team leadership in which one dedicated Scrum Master and an agile team share leadership roles \cite{Moe2010, spiegler2019, Srivastava.2017}. Most studies have examined the Scrum Master role applying qualitative methods \cite{Moe2010, spiegler2019, Srivastava.2017, backlander2019}, while there is a lack in  studies to explore these roles and to understand how much they change \cite{spiegler2019}.

 Studies have found that a Scrum Master influences the ability of a team to work in an agile manner \cite{Moe2010, Gren.2017, spiegler2019, backlander2019, Srivastava.2017}. The Scrum 
 Master role not only facilitates the Scrum Method but also protects the team from inappropriate external requests and empowers the team to work self-organized and cross-functional towards a common goal \cite{cockburn2001agile}. Different qualitative studies suggest that the Scrum Master role changes while the team matures and that some aspects of it are transferred to team members \cite{Moe2010, Srivastava.2017, spiegler2019}. While some studies suggest that the Scrum Master role is entirely transferred to developers in more mature teams \cite{Srivastava.2017, backlander2019}, other studies find that one dedicated Scrum Master plays the role differently in more
mature teams \cite{Moe2010, gren2019agile, spiegler2019}. For example, the Scrum Master is assumed to evolve from command-and-control behavior to a coach \cite{Moe2010, gren2019agile}.
Yet, further understanding of the changing Scrum Master role in relation to maturity is needed \cite{gren2019agile, spiegler2019}. For example, we still lack in quantitative support for a mature team predominantly playing the Scrum Master activities \cite{spiegler2019}.
  
The body of knowledge indicates a need to start quantifying such complex constructs. With this paper, we aim to contribute to understanding agile teams by expanding knowledge on the leadership role of the Scrum Master.
A former study by Spiegler et al. \cite{spiegler2019} examined the activities of a Scrum Master by applying Grounded Theory and identified nine leadership roles which, for reasons of brevity, we label the \textit{9-Factor Theory of Scrum Master Roles}. 
Among their results, the authors found seven of the nine roles to be transferred from a dedicated Scrum Master to the team while it matured over time. The results of a Grounded Theory study are a new theory for future quantitative work \cite{Glaser.2017}. 

The present study builds on Spiegler et al.'s \cite{spiegler2019} theory by providing first empirical data on the 9-Factor Theory.
 Through a quantitative exploration, the present study aims to build groundwork on examining leadership in agile teams quantitatively and shed light on the distribution of leadership roles among the Scrum Master and the agile team with respect to team maturity. It is not our aim to test the process of the role transfer from one Scrum Master to the agile team. 

 Our research questions, inspired by \cite{spiegler2019}, are therefore:

\begin{itemize}
   
    \item Which leadership roles does the Scrum Master play? (RQ1)
    
    \item Which leadership roles does the agile team play? (RQ2)
   
    \item Are leadership roles distributed between a Scrum Master and the agile team, and if so, is the role more often shared in mature as compared to immature teams? (RQ3)
    
\end{itemize}

To answer our research questions we designed an online survey, aimed to quantify the presence of the 9 factors and the maturity of the team.
Sixty-seven participants from more than 19 different  Scrum teams at the Robert Bosch GmbH, an international company which is active in the automotive, industrial and consumer industry, took part in the survey. Through descriptive statistics of the collected data, we found that the leadership roles are shared to a varying extent  between one dedicated role keeper and the agile team. While the Diciplinizer on Equal Terms (explained in Section \ref{ssec:factors}) was shared most often, the Method Champion was shared least often.

Moreover, our data support a changing Scrum Master role such that it was shared most often in teams that had been working between 3 to 5 months in the agile manner. Yet, the percentage of teams who did share the roles was only about 20\% and no agile team predominantly played the Scrum Master role.
 We therefore conclude that despite sharing of some of the 9 factors, the role predominantly sticks with one dedicated Scrum Master.
 
 Based on our results, we suggest to group the 9 factors along three different clusters: psychological team factors, organizational factors and product-related factors. While psychological factors were linked most often to the Scrum Master, organizational factors were assigned less often to both parties.

To be able to support organisations in the agile transformation, we provide empirical evidence on  leadership in agile teams.   We conclude with a suggestion for practitioners on the role description of a Scrum Master which can be implemented in real organizational settings.
We suppose our results are valuable input for future quantitative testing of the 9-Factor Theory.

 \section{Related Work}
  In the following we describe team maturity and the distribution of the Scrum Master role in relation to team maturity, ergo the changing Scrum Master role.
  
 \subsection{Team Maturity}
 
    Team literature research differentiates between static and dynamic teamwork models. While the first refers to teams that are stable and have successfully reached a constant mature stage, the second assumes that a team undergoes different maturity stages. 
This study refers to dynamic teamwork models since we believe it helps us in explaining the changing Scrum Master role. 

An agile team transfers through the different maturity stages until it evolves into a truly agile team \cite{Gren.2017} and, therefore, developers practice the agile way of working differently over time.
    Agile teams are linked to the forming-storming-norming-performing model by Tuckman \cite{tuckman1965}, which we now summarize.

The forming phase  suggests that team members focus on a leader who sets ground rules for further cooperation \cite{tuckman1965}. Team members are insecure on how to behave, and they search for opportunities to observe expected behavior. In this stage, agile teams are suggested to be more open towards leadership that is centred on one person \cite{gren2019agile}.
The storming phase often involves role conflicts due to a lack in unity and security \cite{tuckman1965}. Performance often drops in this stage \cite{katzenbach2015wisdom}.
The norming phase helps teams to increasingly understand and agree on how to work in an agile way \cite{gren2019agile} and to build a shared understanding of roles and responsibilities \cite{neuman1999team}. Team performance increases in this phase \cite{katzenbach2015wisdom}.
The performing stage describes a high performing team in which the team members play roles flexibly according to the situation \cite{tuckman1965}.

\subsection{The Changing Scrum Master Role}

 Several authors assume that the Scrum Master role changes depending on the
 maturity of the team \cite{gren2019agile, Srivastava.2017, Moe2010, spiegler2019, backlander2019}.
Moe et al. \cite{Moe2010} report on teamwork challenges of a newly implemented Scrum team over a period of nine month. They observe that initially the team leadership role was rather centred on the Product Owner and the Scrum Master. The Scrum Master even started to control team members which diminished team leadership and led to less motivation and trust of the team. 
While the team matured, the authors observed  that team leadership advanced, such that team members started to take on more responsibility.

Even though several studies find similar results \cite{gren2019agile, spiegler2019, Srivastava.2017, backlander2019}, researchers do not agree on the extent to which the team plays the Scrum Master role over time. 
 While some authors speculate that only some of the Scrum Master activities are transferred to the team \cite{gren2019agile, spiegler2019, Moe2010}, other authors suggest that the dedicated Scrum Master becomes obsolete in more mature teams \cite{backlander2019, Srivastava.2017}.
   While a study by Backländer \cite{backlander2019} describes that often developers grow into the Scrum Master role over time, Moe et al. \cite{Moe2010} discover that team members rarely take over responsibility. Srivastava and Jain \cite{Srivastava.2017} conclude that all team members should be able to take on the Scrum Master role in more mature teams.  
  
Spiegler et al. \cite{spiegler2019} study  suggests a set of 9 leadership roles of which 7 are gradually transferred to the team, while 2 of the roles remain with one dedicated Scrum Master.  
Their discovered roles are  Method Champion, Disciplinizer on Equal Terms, Change Agent, Helicopter, Moderator, Networker, Knowledge Enabler and Protector, which we summarize in Section \ref{ssec:factors} but are explained to a greater extent in their paper.
We name the nine leadership roles of a Scrum Master the 9-Factor Theory in the present paper. 
 
Since the  Spiegler et al. \cite{spiegler2019} study is a Grounded Theory based theory grounded in empirical qualitative data, the 9 leadership roles of a Scrum Master and how the role distribution unfolds in an immature as compared to a mature team has not yet been quantitatively analyzed.

\section{Method}

This section portrays the participants, the measurement, data collection and analysis of our study.

\subsection{Company Context and Participants}
\label{ssec:company}   
Our data was collected from the multi-national conglomerate Robert Bosch GmbH with more than 20 different sub-companies producing automotive, electrical and consumer industry goods. Scrum teams have the roles Product Owner, Scrum Master and agile team.  Depending on the setting teams may have additional roles like a project manager, business owner, group leader or release train engineer.  Yet, there is no company-wide standard.

The Scrum Master is a job title at the Robert Bosch GmbH. The person playing the committed Scrum Master varies among teams. For example, the role keeper can be a developer or a former group leader. 
Often, the Scrum Master is called 'Agile Master' indicating that the role keeper should rather focus on team dynamics than on the Scrum method. Scrum Masters at the Robert Bosch GmbH are usually not disciplinary supervisors of agile team members, and were probably without authoritative power in our sample. 
 
In total, \textit{67 participants} took part in our study. 46 were from software development projects, 3 from software and hardware development, 4 from software development and IT and the remaining 14 from other topics (e.g.\ mechanical engineering, purchasing, human resources).   56.7\% of the participants had been working more than 11 months with their colleagues.

Our sample contained \textit{37 Scrum Masters} of which 20 had at least 10 months of experience in the Scrum Master role. 
The remaining \textit{30 participants were team members}. 14 team members stated that they were 9 or more members in their team. We did not measure this item for the Scrum Masters.

       Due to confidentiality reasons, providing the team name was optional. 37 participants opted to enter their team name and related to \textit{19 different teams from nine different business divisions} at the Robert Bosch GmbH.
   Since not all respondents inserted their team name, we could not map responses to teams and were only able to compare individual responses.
   
\subsection{Measurement}
The research questions  guiding this study required a quantitative exploration of Spiegler et al.'s 9-Factor theory \cite{spiegler2019}. Each of the nine factors describes a leadership role. 
Besides evaluating the existence of different leadership roles, this study aimed at providing evidence that leadership roles are shared between a Scrum Master and an agile team and that the leadership roles are distributed differently depending on the maturity of an agile team. 

We now briefly describe the 9 Factors. A deeper description is offered in the introductory paper \cite{spiegler2019}.

\textit{Factor MC (Method Champion): The role contains organizing meetings, teaching the method, support formulating tasks and setting goals, and discusses how to adapt the method during the retrospective.}
     
\textit{Factor DE (Disciplinizer on Equal Terms): Supports the team to keep to the rules, ensures that the team focuses on relevant topics and makes sure that team members attend the meetings. Discipline is accomplished via communication on a par.}
    
\textit{Factor CO (Coach): Observes team members and uncovers which kind of behaviour is missing in a team to improve teamwork, provides feedback, and helps teams to find out what they wish to change and how to do so.}

\textit{Factor CA (Change Agent): Serves as a role model, changes habits, and convinces newly established project teams of the agile way of working.}

\textit{Factor HEL (Helicopter): Possesses the ability to see the bigger picture, to know who possess the right skill for a certain task, to include relevant stakeholders and to structure work.}

\textit{Factor MO (Moderator): Moderates all kind of meetings and builds a bridge between perspectives and domains.}

\textit{Factor NET (Networker): Connects the team with relevant stakeholders from within and outside the organisation.}

\textit{Factor KE (Knowledge Enabler): Realises which kind of knowledge the team needs, supports team members to acquire that knowledge and promotes iterative learning.}

\textit{Factor PRO (Protector): Shelters teams from inappropriate requests from the Product Owner, managers, disciplinary leaders and other departments.}

\textbf{Items for Measuring the 9 Factors\label{ssec:factors}}
Based on the description of the Scrum Master roles by Spiegler et al. \cite{spiegler2019}, we initially built a set of 67 items.
Based on techniques rooted in pool items and item review \cite{rust2009}, after two revisions we reduced the initial set to 55 items, each connected to one activity of the nine different roles. 

Each factor was covered by 4 to 9 different items. For example, the Disciplinizer on Equal Terms contained the following four items: \textit{Supports team to keep to the rules. Helps team to focus on relevant topics. Makes sure members attend meetings. Communicates on a par.}
Yet, items are not grouped in the questionnaires, s.t.\ participants are blind to the existence of the factors.  This helps avoid bias that could artificially form clusters.

\textbf{Maturity}
To test maturity, we asked how many months the team had been working in an agile manner. The choice is inspired by Wheelan et al. \cite{wheelan2003group}. They found a significant correlation between the average number of months a team had been working together and the four group development stages \cite{wheelan1996validation}, in which a mature team was perceived to be meeting 5.2 months or more on average (Stage 3=5.2 months on average; Stage 4=8.5 months on average). Based on previous results the question \textit{How many months has your team been working in an agile manner?} provided five choices (0-2 months, 3-5 months, 6-8 months, 9-11 months, more than 11 months).

\begin{table}
  \caption{Maturity}
  \label{tab:freq-maturity}
  
  \begin{center}

  \begin{tabular}{ccc}
    \hline
    Months&Team Member&Scrum Master\\
   &(N=29)&(N=36)\\
    \hline
  0-2 & 0 & 2\\
    3-5&5& 8\\
   6-8& 6& 3\\
   9-11& 4& 5\\    
More than 11& 14& 18\\
  \hline
\end{tabular}
\end{center}

\end{table}

\textbf{Self-Assessment and External Assessment}
Since teams and formal leaders often rate leadership behavior differently \cite{crevani2010leadership}, we conducted a self-assessment and an external assessment for evaluation of each item (leadership activity). Therefore, each item contained two Likert items: the self-assessment and the external assessment. More specifically, the Scrum Master conducted a self-assessment of the leadership behavior he or she believed to perform and an external assessment of the leadership activities he or she believed the agile team performed, and the agile team vice verse rated itself and the Scrum Master.

Therefore, the participants answered each item twice (2*55): one to rate the Scrum Master and one to rate the team. 
The participants rated their perception of leadership activities displayed by the Scrum Master and the agile team using a five-point Likert item with
1=strong disagreement that the activity was done by the respective party, 5=high agreement, and an additional option = Don't know/Not applicable. Questions were randomly ordered.

\subsection{Data Collection}

To assess the 9-Factor Theory we used a web-based survey tool provided by the Robert Bosch GmbH, as part of the agreement to run the study with them.

To invite Scrum practitioners to take part in our survey, we used our personal network within the Robert Bosch GmbH and a internal social business platform provided by the company. An invitation letter contained the link to the online survey and introduced the broader topic of the research and informed that data would be treated anonymously and that participation was voluntarily. 
Besides treating personal data confidentially on our side, participants had the opportunity to voluntarily insert their team name and their email address to receive their aggregated team results. 
This personal data was used for the respective team retrospective only and for no scientific or management purpose, which was also emphasized in the invitation letter. Filling out the survey took approximately 15 minutes. With the exception of the personal data all questions were compulsory.
The full questionnaire is available online \cite{spiegler2020}. Due to confidentiality requirements by the Robert Bosch GmbH, the raw data cannot be provided openly.

\subsection{Pilot study}

    Eight individuals filled out a pilot of the online survey and provided feedback on understanding the content of the items and the convenience to answer the survey. 
  
  Some participants had stated to be annoyed when they had to read one item twice on consecutive pages separately for the Scrum Master and the agile team and the company had urged to build a questionnaire that would not take longer than 15 minutes to be filled out. Rating each item for both parties at the same time and on one page was considered to save time and to be more convenient.

    Even though we had used the feedback for modification, drop out rate was 60\% after launching the survey officially. Several participants delivered the feedback that reading all the items on one page was inconvenient. Therefore, we modified the questionnaire once again, and put the 55 items on three consecutive pages each containing an equal number of items.
    
    This modification led to a loss of data, which we could not plan for with the tool supplied by the company, in 8 already fully filled-out responses. The modified survey accomplished 121 responses, of which 68 were completed while 53 did not reach the last item. We opted to retain only fully completed questionnaires rather than adding partial data.
      16 respondents stopped after they had filled out the first block of items, while 22 respondents dropped out when reaching the first block of items and 15 individuals just opened the link without answering any of the questions.  Once again, we received the feedback by participants, that the questionnaire was inconvenient to be read.
      
      Due to the above-mentioned constraints we still kept the questionnaire the way it was designed. Also we cannot say with certainty why so many individuals decided to stop filling out the questionnaire. It may also be that they did not feel comfortable with rating Scrum Master and agile teams separately.
      
    We removed the responses of one individual who rated every item with ``agree,'' likely indicating a lack of motivation to participate in the study. This led to a total sample of 67 (55.37\%) respondents.

   \subsection{Analysis}
\label{ssec:analysis}   

    For each of the 9 factors we build a mean value by the related items for the Scrum Master and the agile team separately. 
    To avoid including individuals that had only answered a few items  related to one factor, we included responses in the calculation of the mean value when individuals  had at least answered n-1 items per role. That means, if a factor had 4 items, we only included individuals that had answered at least 3 of the items.

To assess whether leadership roles were shared between the Scrum Master and the agile team we applied a similar approach as Zafft, Adams and Smith's \cite{zafft2009measuring} approach to measuring leadership distribution in self-managed teams. Applying a 5-point Likert scale (1=strongly disagree, 5=strongly agree), they suggest a leadership behavior to be present when someone scores higher than 4.0 \cite{zafft2009measuring}. In our analysis, we considered a factor to be embodied by the Scrum Master or the team if the respective party rated 4.0 or higher. If one participant rated both, Scrum Master and agile team, in one factor higher than 4.0, the respective role was considered to be distributed between both parties within one team.

If at least five of the nine factors were found to be shared within the same team, we considered the Scrum Master role to be shared between the agile team and the dedicated Scrum Master.

\section{Results}

 The results are structured as follows: After referring to external and self-assessment, we will answer our three research questions in consecutive order. 

\textit{External and Self-Assessment}
   The average mean for the nine factors revealed that the Scrum Master tended to rate herself higher than the team rated the respective Scrum Master, 
   while the Scrum Master tended to rate the agile team lower. 
   One exception was the Networker which the Scrum Master rated slightly higher than the team rated itself. 
    Likewise, we found that the team members tended to rate themselves higher than the Scrum Master rated them, 
    while they tended to rate the activities performed by the Scrum Master lower.

\subsection{Scrum Master}

  Our first research question is:     \textit{Which leadership roles does the Scrum Master play?} (RQ1)

    To be able to give evidence on the Scrum Master performing one of the nine leadership roles, the mean value of a factor has to be higher than 4.0 (explained in Section \ref{ssec:analysis}). The mean value for four factors is  higher than 4.0, namely Factor MC, DE, CO and MO, and more than two third of the Scrum Masters score high on them. 
    Factor CA, HEL, NET and PRO are linked to about half of the Scrum Masters. Only about one third have a mean value higher than 4.0 regarding Factor KE. 
    More information in Table 2.

\begin{table}
  \caption{Descriptive Statistics for the 9 Factors}
  \label{tab:freq-desc-stat}
  \begin{tabular}{c|ccccc ||ccccc }
    \hline
       &  & \centering{ Scrum  Master} & &  &  & & \centering{Agile Team} & \\
      \hline
   Factor& N & Mean & Std. &  n* &  h**     & N & Mean & Std. &  n* & h** \\
    & & & deviation    & & &&& deviation&\\
    \hline
    MC& 67 & 4.15 & .56 &  47 & 70.15\%    & 60 & 3.19 & .67 & 7 & 11.67\% \\
     DE & 67 & 4.18 & .55 &  49  &73.13\%   & 65 & 3.83 & .52  &  32 &  49.23\% \\
    CO& 66 & 4.09 & .73 &  46 & 69.69\%     & 64 & 3.58 & .59 &  17 &  26.56\% \\
    CA& 61 & 3.95 & .65 &  37 & 60.66\%     & 56 & 3.56 & .52 & 16 & 28.57\% \\
     HEL& 64 & 3.73 & .68 &  28 &  43.75\%   & 62 & 3.72 & .54 & 24  & 38.71\% \\
    MO& 67 & 4.07 & .63 &  49 &  73.13\%    & 62 & 3.72 & .48 & 22 &  35.48\% \\
    NET& 65 & 3.70 & .86 &  30  &  46.15\%   & 62 & 3.44 & .81 &  21  &  33.87\%  \\
    KE& 63 & 3.62 & .76 &  22  &  34.92\%   & 58 & 3.58 & .66 &  20  & 34.48\%  \\
     PRO& 62 & 3.70 & .88 &  32  & 51.61\%   & 53 & 3.10 & .84 & 10 & 18.86\% \\

  \hline
\end{tabular}
*n describes the absolute frequency of a factor rating higher than 4.0.\ 
**h describes the relative frequency (n/N per row). 

Note: Each column contains summarized results and refers to answers by Scrum Masters and the agile team taken together.
\end{table}

     Therefore, we answer RQ1 and find that a majority of the Scrum Masters play the Method Champion, Disciplinizer on Equal Terms, Coach and Moderator, while the Change Agent, Helicopter, Networker and Protector is played by merely about half of the Scrum Masters and the Knowledge Enabler is performed by only about one third.

\subsection{Agile Team}

    Our second research question is: \textit{Which leadership roles does the agile team play?} (RQ2)

    To be able to give evidence on the team playing one of the nine roles, the mean value of a factor has to be higher than 4.0 (explained in Section \ref{ssec:analysis}).
Table 2 illustrates that all mean values  of the nine factors related to the agile team are lower than 4.0. 
  Therefore, one could claim that team members tend to not play the leadership roles.
    Yet,  almost 50\% of the teams score higher than 4.0 for Factor DE. Between 30\% and 40\% perform Factor HEL, MO, NET and KE. Factor MC and PRO are rarely aligned to the team.

        Based on our results, we answer RQ2 and find that the agile team tends to not play the leadership roles. About half of the teams perform the Disciplinizer on Equal Terms, while only about one third perform the Helicopter, Moderator, Networker and Knowledge Enabler. The Method Champion, Coach and Protector are performed least often by the teams.

\subsection{Distribution of the 9 Factors between Scrum Master and Agile Team}

The third research question is: \textit{Are leadership roles distributed between a Scrum Master and the agile team, and if so, is the role more often shared in mature as compared to immature teams?}

   If a participant scores  a factor for both the Scrum Master and the team higher than a mean value of 4.0, the factor is considered to be distributed between the Scrum Master and the agile team. 
   While Factor DE, HEL and MO are distributed in 30\% to 40\% of the teams, Factors MC, CA, KE and PRO are distributed in 10\% to 20\% of the teams.
    Table 5 shows an overview on the distribution for each of the nine factors, starting with the most frequently shared Factor DE to the least frequently shared Factor MC.

  \begin{table}

  \label{tab:freq-distribut}  
  
  \caption{Distribution of the 9 Factors}
  \begin{center}
  \begin{tabular}{ccccc|cc}

  \hline
  
  \toprule
  Factor& \enspace Shared & \enspace Only Scrum Master & \enspace Only Team & \enspace No one & \enspace N & \enspace Total \% \\

   \hline
   
    DE & 43.30\% & 29.90\% & 4.50\% & 22.40\% & 67 & 100.00\% \\
  
    MO & 31.30\% & 41.80\% & 1.50\% & 25.40\% & 67 & 100.00\% \\
    
    HEL & 28.40\% & 13.40\% & 7.50\% & 50.70\% & 67 & 100.00\% \\

    CO & 25.40\% & 43.30\% & 0.00\% & 31.30\% & 67 & 100.00\% \\
    
   NET  & 22.40\% & 22.40\% & 9.00\% & 46.30\% & 67 & 100.00\% \\
   
    CA & 19.40\% & 35.80\% & 4.50\% & 40.30\% & 67 & 100.00\% \\
    
     KE & 16.40\% & 16.40\% & 13.40\% & 53.70\% & 67 & 100.00\% \\
    PRO & 14.90\% & 32.80\% & 0.00\% & 52.20\% & 67 & 100.00\% \\
    
    MC & 10.40\%  & 59.70\% & 0.00\% & 29.90\% & 67 & 100.00\%\\
    
   \hline

\end{tabular}
\end{center}

  Note: Each column contains summarized results and refers to answers by Scrum Masters and the agile team taken together.

\end{table}

 If a respondent scores a mean value higher than 4.0 for at least five of the factors for both, Scrum Master and agile team,  the Scrum Master role is considered to be distributed between the agile team and the dedicated Scrum Master. 20.90\% of the respondents share the Scrum Master role.

38.5\% of the teams that had been working 3-5 months in an agile manner shared the Scrum Master role, 11.11\% of the teams rating  6-11 months shared it and 18.8\% of the teams rating more than 11 months shared the role. Therefore, teams that had been working for 3-5 months tended to share the role by 20 percentage points more than teams that had been working for 11 months or more, and by 27.39 percentage points more than teams that had been working in an agile way between 6-11 months.

Furthermore, we check if some teams perform the Scrum Master role predominantly, such that the team scored for 5 factors higher than 4.0, while the Scrum Master scored for less than 5 factors higher than 4.0. We did not find such a case in our data.

 Based on these results we answer RQ3 and claim that leadership roles can be shared, yet, some roles are shared more often than others.
While we find that the Disciplinizer on Equal Terms is most often shared between the team and the Scrum Master, we find that the Method Champion, Coach and Protector are rather centred on one dedicated Scrum Master. 
 
 Furthermore, the distribution of the Scrum Master role varies in different maturity stages. 
 We find that teams who share the role had most often been working in an agile way between 3 to 5 months. Therefore, the role was rather shared in immature teams. Furthermore, we did not find a single team in which the Scrum Master role was centred on the agile team.

\section{Discussion}
    \label{ssec:discussion}

     Our study aimed at exploring the presence of and the change in the 9-Factor Theory \cite{spiegler2019}.
     Based on descriptive statistics,  we found that the 9 different roles are performed to a varying extent:
   
   While the Scrum Master rates highest in the Method Champion, Disciplinizer on Equal Terms, Coach and Moderator, the agile team scores highest in the Disciplinizer on Equal Terms, Helicopter, Moderator, Knowledge Enabler and Networker. Both, Scrum Master and agile team, tend to perform the Protector less often than the other roles.
    
     Based on this result, we suggest to broaden the 9-Factor Theory by Spiegler et al. \cite{spiegler2019}. Our results indicate that the 9 factors can be further grouped into three clusters: psychological team factors, product-related factors and organizational factors. We will now elaborate on this idea based on empirical results.
  
  Factor MC, CO and MO rather focus on internal socio-psychological team mechanisms, while Factor CA, NET and PRO involve an external focus towards the organization. Factor DE, HEL and KE are rather product-related and aim at continuous learning and knowledge sharing. The Scrum Master scores higher in roles related to psychological team factors (e.g.\ Method Champion and Coach). The team scores higher in product-related factors (e.g.\ Helicopter and Knowledge Enabler). Roles that bridge the organization with the team were played more often by the agile team regarding the Networker, but less often regarding the Protector.

  \begin{table}
  \caption{3 proposed Clusters of the 9-Factor Theory}
  \label{tab:freq}
  
  \begin{center}

  \begin{tabular}{ccc}
    \hline
    Cluster &Leadership Role (Factor) & More important to\\
   
    \hline
 Psychological &  Method Champion (ME) & Scrum\\
 Team Factors & Coach (CO) & Master \\
    &  Moderator (MO)  \\
    \hline
    Product-Related  & Disciplinizer on Equal Terms (DE)& Agile  \\
   Factors &  Helicopter (HEL)  &  Team\\
   & Knowledge Enabler (KE)\\
    
    \hline
    
    Organizational  & Change Agent (CA) & It \\
    Factors & Networker (NET) & depends \\
    &  Protector (PRO)\\
  \hline
\end{tabular}
\end{center}

\end{table}

  Moreover, about half of the teams did not play the Protector, the Change Agent or the Networker which are linked to the organizational factors.
  In rather bureaucratic organizations, as in our case, it might be more difficult to perform the roles related to bridging the organization and the team. A traditional environment rather focuses on hierarchy as opposed to protect the team from management and on departmentalized structure as opposed to network with each other independent from formal structures \cite{nerur2005challenges}.

  We speculate that if a Scrum Master played the Protector to a larger extent, the agile team would take over the leadership roles more often. The Protector provides hierarchical free space within which team members feel safe to take on the divers roles \cite{spiegler2019}.
  
  Furthermore, 53\% of the teams did not perform the Knowledge Enabler and about 51\% the Helicopter.  
  A possible explanation for our results would be that either the Scrum Master considers product-related roles to not be part of the job description since the agile team is expected to self-organize their work, or it is more difficult to play the respective roles in a bureaucratic context since that company type is build on experts with specialized skills as opposed to cross-functional knowledge sharing \cite{nerur2005challenges}.  This may be supported by the teams scoring equally low on this factor.

     This study also aimed at exploring the 9-Factor Theory in relation to maturity. The 20\% of the teams that did share the Scrum Master role, provided support for the suggestion that the Scrum Master role is distributed differently in different maturity stages. Teams that had been working in the agile manner for 3 to 5 months and more than 11 months shared the role most often.

   This finding fits with the maturity model by Tuckman \cite{tuckman1965}:
   Teams after 3 to 5 months tend to be in the storming phase, within which teams are not sure about who plays which role within the team. Therefore, both, team and Scrum Master, perform the Scrum Master roles. Teams working in an agile way for more than 11 months could already have reached the performing phase within which roles are played according to the situation and less linked to one dedicated role keeper.

   Yet, we did not find any agile team that played the Scrum Master role to a larger extent than the dedicated role keeper.  Therefore, our results do not point at the direction that the formal role keeper steps back from the role as suggested  by several studies \cite{backlander2019, Srivastava.2017, spiegler2019}.
    This finding also fits with earlier claims that teams in organizational settings rarely develop into high performing teams that take on roles spontaneously \cite{marks2001temporally}. We therefore propose that in most of the teams the dedicated Scrum Master does not become obsolete over time but rather changes the primary role during the different phases of team development.

   Another explanation of the results could be that neither Scrum Master nor agile team but someone else took over the role. As described in Section \ref{ssec:company} agile teams and the aligned agile roles vary among different settings at the Robert Bosch GmbH. It might be that some of the nine leadership roles are also played by the Product Owner or disciplinary supervisor. However, those roles were neglected in our study. The last paragraph of the practical implications provide suggestions on how to deal with this in company settings.
    
    \section{Practical Implications}
  We found that the leadership roles were rather centred on the Scrum Master.  In the following we thus suggest how to develop the Scrum Master description in company settings.
    Section \ref{ssec:discussion} proposed to group the Scrum Master description into three clusters: psychological team factors, organizational factors and product-related factors. 
    
    While some practitioners suggest that the Scrum Master should play product-related roles, others state that interference on a technical level hinders self-organizing teams. We suggest that every team should discuss on its own, to which degree it needs product-related support by a Scrum Master.
    Yet, a Scrum Master who performs product-related roles builds an understanding of the respective product, thus, can also more easily bridge the agile team with the processes, requirements, tools and standards of a rather bureaucratic surrounding. 
   
    For example, the Scrum Master can be a mouthpiece of the team to discuss with the management which processes and requirements of rather traditional project management are still needed despite the team working in an agile way, and which ones are rather unnecessary and hinder the progress of the team. The Scrum Master can argue which tools and processes the team needs to work in a more agile way. 
    Also, taking over product-related roles improves understanding when to protect the team, e.g.\ from re-prioritization, and when to give in and allow to re-arrange planning due to changes in requirements on organizational level. 
    
    Thus, the Scrum Master supports the organization to gradually evolve into a more agile place. 
    Yet, we acknowledge the balancing act of a Scrum Master to support the team in product-related matters and to serve as a coach at the same time.
   The Scrum Master continuously needs to serve as a coach and support the team to learn how to take on the divers roles. 
    
       Agile teams in a traditional industrial conglomerate may not be used to take on leadership activities as a whole team.  Yet, if the leadership gap \cite{spiegler2019} is not filled by the agile team, there is the risk of a leadership vacuum, in which no one takes over leadership roles. This may lead to less performance. 
 
    Nevertheless, we found that the Scrum Master and the team tend to play organizational factors to a lesser extent, and encourage managers even further to build an agile friendly surrounding within which the organizational factors can be performed. These factors are necessary to integrate the agile team into the organizational setting, such as having access to relevant stakeholders and information, reducing interfaces and efforts for alignment and building trust between agile teams and traditional structures. 
    Consequently, motivation and progress of agile teams will increase even further.
    Yet, organizations also need to understand and accept that sometimes teams do not want to take on leadership roles.

      Therefore, companies should use our questionnaire to reflect upon the role distribution in their specific industry background and organizational environment relevant to their team.  There might be roles beyond the Scrum Master and the agile team that take on the leadership roles. 
      Thus, we suggest to not only focus on leadership sharing among the Scrum Master and the agile team but to broaden the perspective. We propose to use the leadership roles and aligned activities to determine if they are covered by any 'job title' in the setting, which might be the Product Owner or the disciplinary supervisor. After all, the agile way of working is not about establishing a standard regarding which job title plays which leadership role but about making sure that the needs of an agile team are covered in any given situation. Since teams mature and agile settings vary, teams need to find a context-dependent equilibrium of leadership sharing. Therefore, each team has to discuss on its own how to divide leadership activities among each other. Furthermore, since context changes, teams need to discuss regularly upon who takes on which leadership role in a given situation.
      Practitioners will understand respective leadership needs, learn to balance and evolve them, and thus, improve teamwork.
     
\section{Limitations and Future Work}
   In the following we will suggest future topics for research while referring to limitations of this study.

    Objectivity: since  we conducted an online survey, we assume a low level of social response bias. Yet, respondents were allowed to insert their email address for receiving their team results. This could lead to a social response bias in such a way that respondents wanted to rate high in the Scrum Master activities.

    With 67 participants our sample is limited in size and prevented us to perform a psychometric evaluation of the tool, limiting our confidence in the tool validity. A psychometric evaluation of the tool would not be a familiar step in software engineering studies, so we see this as a missed opportunity rather than a limitation. Future studies should aim for a bigger sample size that allows to perform an exploratory factor analysis, thus quantitatively clustering the factors.
    As the theorized 9 factors might be difficult to test psychometrically, we suggest future studies to test the three suggested clusters in Section \ref{ssec:discussion}, thus, allowing for testing agile team behavior along three variables instead of nine.
    
    Since the drop-out rate for this study was quite high, for future studies we suggest, to rate the Scrum Master and the agile team each on separate consecutive pages. Therefore, participants will have to answer six different pages of questions. This will take more time, yet, may lead to a more convenient experience to fill out the questionnaire, and thus, increase the number of responses.
     
    Moreover, even though each business division operates within a different sub-culture and industry context, still all teams were from the same conglomerate. Even if our study is clearly placed as an exploratory one, we want to highlight that we cannot claim our results to be universally applicable.
    We suggest a larger sample drawn from different companies with different industry backgrounds to extend our study in the future.

  Moreover, almost 50\% of the team members stated to be 9 or more persons in their team. We were not able to control for this variable since we had not asked the Scrum Master on their number of team members. Larger groups are found to be less likely to evolve into a mature team \cite{wheelan2003group}. Future testing should take this into account.

Our  data points at an evolving Scrum Master role in relation to maturity. However, maturity was rated by the number of months each team had been working in an agile way. We cannot claim with certainty that the time a team has been working in an agile manner is related to maturity stages. Furthermore, we have not conducted a longitudinal study but compared different teams which had been working a varying amount of time in the agile manner.

Future testing should refer to the maturity stage by Wheelan \cite{wheelan2003group} to examine the 9-Factor Theory for a valid measurement of group maturity, investigate time and group development in relation to varied company types and sizes, as well as in a longitudinal study.

\bibliography{sample-base}

\end{document}